\documentstyle[11pt,newpasp,twoside,epsf]{article}
\def\edcomment#1{\iffalse\marginpar{\raggedright\sl#1\/}\else\relax\fi}
\reversemarginpar

\begin{document}
\title{Color Correlations in (S+S) Binary Galaxies I. The Holmberg Effect.}
\author{Hector Hern\'andez Toledo}
\affil{IA-UNAM - Distrito Federal, Mexico}
\author{Iv\^anio Puerari}
\affil{INAOE - Tonantzintla, Mexico}

\begin{abstract}
 Relying on his photometry of galaxies, Holmberg (1958) found,
more than 40 years ago, that the color indices of paired
galaxies were closely correlated. Our deep broad-band BVRI CCD 
photometry of 45 (S+S) pairs from the Karachentsev (1972) 
catalogue (see Hern\'andez Toledo \& Puerari, this volume), 
and additional (B-V) color information from the literature
(50 extra (S+S) pairs), help us to confirm the effect. This ``Holmberg 
Effect'' has long been remained unverified and not explained yet.
  
\end{abstract}

\section{Introduction}

Hints that galaxy interactions might play an important role 
in inducing multi-wavelength emission enhancement date back 
at least 40 years. One of the first quantitative indications 
came from Holmberg (1958), as a byproduct of his famous 
photometric survey of nearby galaxies. By comparing the 
photographic colors of paired galaxies, he found a significant 
correlation between the colors of pair components. This 
phenomenon has since been referred to as the ``Holmberg effect''.
Although the physical explanation of the Holmberg effect 
is complex, it reflects in part a tendency for similar 
types of galaxies to form together (morphological concordance),
a possible reflection of the role of local environment in 
determining galaxy morphology, but it can presumably also 
reflect mutually induced star formation (Kennicutt 1998).

\section{Preliminary Results: The Holmberg Effect}

During 1999, a sample of 45 spiral--spiral (S+S) binary 
galaxies was observed at Observatorio Guillermo Haro, 
Cananea, Sonora,  M\'exico, as a part of our observational 
project on binary galaxies selected from the Karachentsev 
Catalogue (1972).  A basic surface photometry analysis of 
this sample is also presented in this volume 
(Hern\'andez Toledo \& Puerari). The estimated (B-V) 
colors for (S+S) pairs were corrected for both 
internal and galactic interstellar reddening and 
subsequently reduced to the RC3 system. Figure 1 
shows the (B-V) color correlation. The vertical axis 
refers to the brighter (primary) component and the 
horizontal axis refers to the fainter (secondary) 
component. Additional information on colors for 
Karachentsev (S+S) pairs, mainly from the Lyon--Meudon 
Extragalactic Database (LEDA) and the catalogue of 
aperture photometry in UBVRI by Prugniel 
\& H\'eraudeau (1998) were included. Demin et 
al. (1983), M\'arquez \& Moles (1996), and Junqueira 
et al. (1998) also provided some color information.

\begin{figure}  
\includegraphics{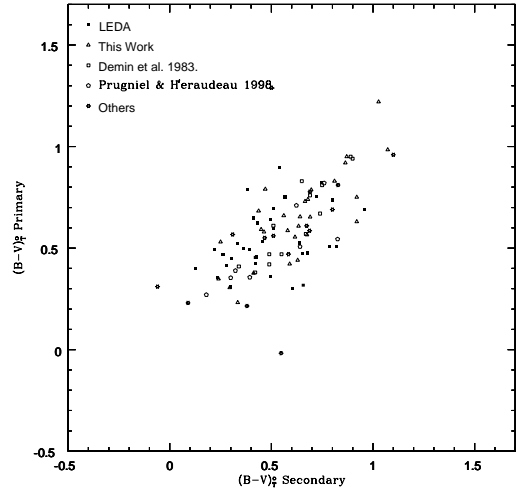}
\vskip6.5truecm
\caption{Correlation between the color indices
(B-V) for the components of (S+S) pairs. The
ordinate shows the color of the brighter component.}
\end{figure}

We can see that the two colors are tightly correlated. A
correlation coefficient $r \sim 0.77 \pm 0.03$ is obtained
for our (S+S) pairs and the additional color information
displayed strongly supports this result. A more careful 
analysis of the (B-V), (B-R) and (B-I) colors involving 
stellar population synthesis models is in progress.

\acknowledgments
HHT and IP were fortunate enough to have
participated in this meeting and want to 
thank the organizers.

\end{document}